# Casimir-Polder attraction-repulsion crossover criterion

Partha Goswami*

Deshbandhu College, University of Delhi, Kalkaji, New Delhi-110019, India
*<physicsgoswami@gmail.com>

**Abstract**
The mutual electromagnetic correlations between two spatially separated systems gives rise to Casimir and Casimir-Polder effect. The corresponding forces, which are generally attractive for most vacuum-separated metallic or dielectric geometries, are due to the contribution to the ground-state energy of the coupled system. We investigate here the Casimir-Polder free energy corresponding to interactions of a magnetically and electrically polarizable micro-particle with a magneto-dielectric sheet. Our semi-phenomenological study shows that such an interaction is reversibly tunable in strength and sign.The latter, particularly, is true provided we look for the exotic materials fabricated at scales between the micron and the nanometer. The crossover between attractive and repulsive behavior is found to depend on the polarizability ratio of the micro-particle and the electromagnetic impedance of the magneto-dielectric sheet.

**MAIN TEXT**
The Casimir/Casimir-Polder effect is attractive in most vacuum-separated metallic or dielectric geometries. The effect is due to the contribution to the ground-state [1,2] energy of the coupled system. Two electrically neutral spatially separated systems interacting via Casimir force will have access to the stable separation state only when the force gets transformed from repulsion at small separations to attraction at large separations. Such issues are important in the future development of micro- and nano-electromechanical systems (MEMS and NEMS).The repulsive Casimir forces [3]are believed to occur in four types of materials, viz. the fluid-separated dielectrics [4], the composite meta-materials [5], the systems with different geometries [6,7], and the time-reversal symmetry (TRS)broken systems [8,9]. It is well-known[10,11.12] that the Weyl semimetal state must break TRS or the inversion symmetry. Therefore, the corresponding host materials are expected to yield a Casimir/Casimir-Polder (CP) repulsion tunable with carrier doping or a magnetic field [13]. Experimentally, the forces have been realized for the first time involving test bodies immersed in a liquid medium- ethanol[4]. We investigate here the Casimir-Polder free energy corresponding to interactions of an electrically and magnetically polarizable micro-particle with a magneto-dielectric sheet. Our task is to look for the repulsive Casimir-Polder forces between a micro-particle possessing non trivial ratio of the magnetic polarizability and the electric polarizability and the artificially engineered dielectric material sheet having non trivial magnetic permeability values. The natural materials have a magnetic permeability roughly equal to one in the range of frequencies relevant for the Casimir effect. We show that for the non-trivial permeability values, the crossover between attractive and repulsive behavior depends on 'polarizability ratio' of the micro-particle, and the impedance $Z = \sqrt{(\mu/\varepsilon)}$ of the sheet apart from the ratio of the film thickness and the micro-particle separation ($D/d$) and temperature($T$). The importance of CP repulsion cannot be understated. The repulsion stabilizes the operation of MEMS and NEMS, as it liberates one from the badgering problem of 'stiction' in such systems.

The Casimir (CI)and Casimir-Polder (CPI)interactions(whereas CI refers to force between two bulk objects, such as dielectric plates, CPI describes the force between a bulk object and a gas-phase atom) are conservative and arise due to the quantum fluctuations of the

photon(electromagnetic) field or, more generally, from the zero-point energy of materials [1,2] and their dependence on the boundary conditions of the photon fields. The Casimir effect due to the vacuum fluctuations of the phonon field, similar to that due to the vacuum fluctuations of photon field, was also predicted for a Bose-Einstein condensate (BEC). A unified picture of these quantum-mechanical, fluctuation-driven-forces was given by Lifshitz [16,17,18] many decades ago: We consider a micro-particle in vacuum where the former is characterized by the dynamic electric polarizability $\xi_{microparticle}(\omega)$ and the dynamic magnetic polarizability $\eta_{microparticle}(\omega)$ as shown in Figure 1. The sample in the figure consists of a thin film of thickness *'D'* deposited on a thick substrate at temperature *T*. Suppose the particle is at a separation *'d'* ($d >> d(T) \equiv \hbar c/(2k_B T)$ or, $T >> T_c \equiv \hbar c/(2dk_B)$) above the sample. Our aim is to investigate the interaction of this micro-particle with the given film material. To elucidate ab initio the concept of the classical Casimir-Polder interaction (CPI), we assume the interaction of the particle with the sheet to be of the *classical* CPI type in the first approximation. The classical case is valid when the separation is not small. The quantities $d(T)$ and $T_c$ set the classical limit in the sense that the limit starts from $d \approx 5\, d(T)$ and $T \approx 5T_c$. To explain, for ordinary materials at room temperature (300 K), $d(T) = \hbar c/(2k_B T)) \approx 3.66$ μm. Thus, the classical limit is achieved in this case approximately at separations above the edge $d = 10$ μm. In terms of temperature, for the separation $d = 10$ μm, the classical limit edge is $T_c \approx 110$ K. Thus, at $T >> 110$ K, the classical limit is nearly achieved. The projection of the wave vector on the $(x,y)$ plane is denoted by $\mathbf{k}^{\perp}$, and $k^{\perp} = |\mathbf{k}^{\perp}|$. We choose the coordinate plane $(x, y)$ coinciding with the upper film surface and the z-axis perpendicular to it (Figure 1). The quantities $\varepsilon^{(0)}(i\omega_l) \to \varepsilon^{(0)}_{intervening\ medim}(i\omega_l)$ and $\mu^{(0)}(i\omega_l) \to \mu^{(0)}_{intervening\ medium}(i\omega_l)$ are the dynamic dielectric (relative) permittivity and the dynamic magnetic (relative) permeability of the intervening medium. If the medium happens to be vacuum and then static counterpart of each of them is equal to one. The micro-particle is characterized by the static polarizabilities, such as the electric polarizability $\xi_{microparticle}(0)$ and the magnetic polarizability $\eta_{microparticle}(0)$. The (electric) dipole polarizabilities are given in a variety of units, depending on the context in which they are determined. The most widely used unit for theoretical atomic physics is atomic units (a.u.), in which, e, $m_e$, the reduced Planck constant ℏ, and $4\pi\varepsilon_0$ have the numerical value 1. The quantity $4\pi\varepsilon_0 = 1$ implies that the electric polarizability in *a.u.* has the dimension of volume. However, the SI unit is C-m$^2$/V. The magnetic polarizability tensor($\beta_{ij}$) relates the induced magnetic dipole moment $m_i$ to the inducing field $B_j$ according to the equation $m_i = \beta_{ij} B_j$. The magnetic polarizability is also defined by the spin interactions of nucleons. Both the definitions lead to the SI unit C$^2$-m$^2$- kg$^{-1}$. We introduce now the polarizabilities $(\eta(0), \xi(0))$ in the atomic units. It may be mentioned that the polarizability ratio $Z_{2,\ microparticle} = (\eta_{microparticle} / \xi_{microparticle})^{1/2}$, though, has the SI unit m-s$^{-1}$, its dimensionless counterpart is $Z_2 = (\eta(0) / \xi(0))^{1/2}$ in a.u.. We shall use the polarizabilities in a.u. below in defining the free energy. This choice has the great advantage of not have to worry about the dimensions of the various quantities involved in the numerics and the graphics to follow. Suppose the film material is characterized by the dielectric (relative) permittivity $\varepsilon^{(1)}(\omega)$ and the magnetic (relative) permeability $\mu^{(1)}(\omega)$, and the substrate is by the relative permittivity $\varepsilon^{(2)}(\omega)$ and the relative permeability $\mu^{(2)}(\omega)$. These might be made of either dielectric or metallic materials or poor conductor. We further assume that for the film material there exist finite limiting values of the relative permittivity and the permeability: $\varepsilon^{(1)}(0) \equiv \varepsilon_0^{(1)}$ and $\mu^{(1)}(0) \equiv \mu_0^{(1)}$. Correspondingly, the dimensionless impedance of the film material is $Z_1 = (\mu_0^{(1)} / \varepsilon_0^{(1)})^{1/2}$. If the film happens to be on a substrate then the static counterparts of $\varepsilon^{(2)}(i\omega_l)$ and $\mu^{(2)}(i\omega_l)$ are not equal to 1; the quantities $\varepsilon^{(2)}(i\omega_l)$ and $\mu^{(2)}(i\omega_l)$ are the dynamic dielectric permittivity and the magnetic permeability of the substrate material, respectively. Suppose, the finite limiting values of the static counterparts these quantities are $\varepsilon_0^{(2)}$ and $\mu_0^{(2)}$. If the film happens to be isolated, then $\varepsilon_0^{(2)}$ and $\mu_0^{(2)} = 1$.

We denote the reflection coefficients of the electromagnetic fluctuations on the sheet material plus substrate, dependent on the wave vector projection $k^\perp$ on the $(x,y)$ plane (and also on the frequency), for two independent modes, viz. the transverse magnetic (TM) and the transverse electric (TE) polarizations, by $\Pi^{(0)}{}_m(i\omega_l, k^\perp)$, and $\Pi^{(0)}_e(i\omega_l, k^\perp)$, respectively. Here $\omega_l = (2\pi l k_B T/\hbar)$ are (imaginary) Matsubara frequencies. The dependence on '$\omega_l$' is borne out by the fact that the Casimir/ Casimir-Polder force not only arises from the fluctuations of the electromagnetic field, which are purely quantum-mechanical objects, they also have a thermal contribution [1,18] at nonzero temperatures. The closed-form, precise expressions for these reflection coefficients, including the thermal contribution, are given by the Fresnel coefficients $\mathfrak{I}^{(n,n')}{}_m(i\omega_l, k^\perp)$, and $\mathfrak{I}^{(n,n')}{}_e(i\omega_l, k^\perp)$ ([15,16,17]) (the Fresnel coefficients are calculated along the imaginary axis ([15,16,17]) corresponding to the reflection on the boundary planes between the vacuum and the film material $(n = 0, n' = 1)$ and also between the film material and the substrate $(n=1, n'=2)$:

$$\Pi_\alpha^{(n=0\to n'=1),(n=1\to n'=2)}(i\omega_l, k^\perp) = \frac{(\mathfrak{I}^{(0,1)}{}_\alpha(i\omega_l,k^\perp) + \mathfrak{I}^{(1,2)}{}_\alpha(i\omega_l,k^\perp)\exp(-2k^\perp D))}{(1+ \mathfrak{I}^{(0,1)}{}_\alpha(i\omega_l,k^\perp)\mathfrak{I}^{(1,2)}{}_\alpha(i\omega_l,k^\perp)\exp(-2k^\perp D))}, \alpha=(m,e). \quad (1)$$

Here, the indices $\alpha,\beta=(m, e)$ denote the transverse magnetic (TM) and the transverse electric (TE) modes. The Fresnel coefficients, which describe the reflection and transmission of electromagnetic waves at an interface, are given by

$$\mathfrak{I}_m{}^{(n,n')}(i\omega_l, k^\perp) = \frac{\left[\varepsilon^{(n')}(i\omega_l)k^{(n)}(i\omega_l,k^\perp) - \varepsilon^{(n)}(i\omega_l)k^{(n')}(i\omega_l,k^\perp)\right]}{\left[\varepsilon^{(n')}(i\omega_l)k^{(n)}(i\omega_l,k^\perp) + \varepsilon^{(n)}(i\omega_l)k^{(n')}(i\omega_l,k^\perp)\right]}, \quad (2)$$

$$\mathfrak{I}_e{}^{(n,n')}(i\omega_l, k^\perp) = \frac{\left[\mu^{(n')}(i\omega_l)k^{(n)}(i\omega_l,k^\perp) - \mu^{(n)}(i\omega_l)k^{(n')}(i\omega_l,k^\perp)\right]}{\left[\mu^{(n')}(i\omega_l)k^{(n)}(i\omega_l,k^\perp) + \mu^{(n)}(i\omega_l)k^{(n')}(i\omega_l,k^\perp)\right]}, \quad (3)$$

where $k^{(n)}(i\omega_l, k^\perp) = [k^{\perp 2} + \varepsilon^{(n)}(i\omega_l)\mu^{(n)}(i\omega_l)(\omega_l^2/c^2)]^{1/2}$. The reflection coefficients of the electromagnetic fluctuations on the sheet material plus substrate, viz. $\Pi_m(i\omega_l, k^\perp)$ $(n = 0 \to n' = 1), (n = 1 \to n' = 2)$, in view of Eqs.(2)-(5), for the transverse magnetic (TM) polarization, could be written as

$$\Pi_m{}^{(n=0\to n'=1),(n=1\to n'=2)}(i\omega_l, k^\perp = \kappa/2d)$$

$$= \frac{[\tan(\frac{\pi}{4} - \varphi^{(m)}{}_{0,1}(i\omega_l,\frac{\kappa}{2d})) + \tan(\frac{\pi}{4} - \varphi^{(m)}{}_{1,2}(i\omega_l,\frac{\kappa}{2d}))\exp(-kD/d)]}{[1+\tan(\frac{\pi}{4} - \varphi^{(m)}{}_{0,1}(i\omega_l,\frac{\kappa}{2d}))\tan(\frac{\pi}{4} - \varphi^{(m)}{}_{1,2}(i\omega_l,\frac{\kappa}{2d}))\exp(-kD/d)]}. \quad (7)$$

For the transverse electric (TE) polarization, one may similarly write

$$\Pi_e{}^{(n=0\to n'=1),(n=1\to n'=2)}(i\omega_l, k^\perp = \kappa/2d)$$

$$= \frac{[\tan(\frac{\pi}{4} - \varphi^{(e)}{}_{0,1}(i\omega_l,\frac{\kappa}{2d})) + \tan(\frac{\pi}{4} - \varphi^{(e)}{}_{1,2}(i\omega_l,\frac{\kappa}{2d}))\exp(-kD/d)]}{[1+\tan(\frac{\pi}{4} - \varphi^{(e)}{}_{0,1}(i\omega_l,\frac{\kappa}{2d}))\tan(\frac{\pi}{4} - \varphi^{(e)}{}_{1,2}(i\omega_l,\frac{\kappa}{2d}))\exp(-kD/d)]}. \quad (8)$$

where we have introduced a dimensionless variable in place of $k^\perp$ above, viz. **$\kappa \equiv 2d\,k$ and**

$$\varphi^{(m)}{}_{n,n'}(i\omega_l, k^\perp) = \arctan[(\varepsilon^{(n)}(i\omega_l) k^{(n')}(i\omega_l, k^\perp))/(\varepsilon^{(n')}i\omega_l) k^{(n)}(i\omega_l, k^\perp))], \qquad (4)$$

$$\varphi^{(e)}{}_{n,n'}(i\omega_l, k^\perp) = \arctan[(\mu^{(n)}(i\omega_l) k^{(n')}(i\omega_l, k^\perp))/(\mu^{(n')}i\omega_l) k^{(n)}(i\omega_l, k^\perp))], \qquad (5)$$

In view of Eqs. (7) and (8), one obtains the simple expression for the Casimir-Polder free energy density as

$$F(d) = -\left(\frac{k_B T}{8d^3}\right) \sum_{\substack{\alpha,\beta \\ \alpha \neq \beta}} \sum_l \int_0^\infty \kappa^2 d\kappa \, e^{-\kappa} \{\eta_\alpha(i\omega_l) \Pi_\beta^{(n=0 \to n'=1),(n=1 \to n'=2)}(i\omega_l, \kappa/2d)\}. \qquad (9)$$

The dielectric constant $\varepsilon^{(n)}(i\omega_l)$, the electric polarizability $\eta_e(i\omega_l)$, etc. though written as function of frequency above, in general, is a function of frequency and the wavevector both. They descibe the response of a medium to any field. As alrady mentioned, for the fields slowly varying in space and time, the limiting value of these functions are the Faraday-Maxwell dielectric constant and the static electric polarizability $\eta_e(0)$. We shall assume the magnetic polarizability of the micro-particle and the permeability of the sheet material having the similar limiting static values below.

The important outcome, of the Faraday-Maxwell (static) limit, is that $k^{(n)}(i\omega_l, k^\perp) = k^\perp$. In view of (2) and (3) one is then able to write

$$\mathfrak{I}_m^{(n=0,n'=1)}(i\omega l, k^\perp) = \mathfrak{I}_m^{(n=0,n'=1)}(0, k^\perp) = \mathfrak{I}_m^{(n=0,n'=1)}(0,0) = \left(\frac{\varepsilon^{(1)} - \varepsilon^{(0)}}{\varepsilon^{(1)} + \varepsilon^{(0)}}\right),$$

$$\mathfrak{I}_m^{(n=0,n'=1)}(i\omega l, k^\perp) = \mathfrak{I}_m^{(n=0,n'=1)}(0, k^\perp) = \mathfrak{I}_m^{(n=0,n'=1)}(0,0) = \left(\frac{\mu^{(1)} - \mu^{(0)}}{\mu^{(1)} + \mu^{(0)}}\right), \qquad (10)$$

and so on. The summation $\sum_l$ in (9) disappears in this static limit. It must be clarified that this limit is not the same as the low-temperature limit where $\omega_l$'s will get closer to each other and at zero temperature all of them contribute to dissipation. Thus, the Casimir-Polder free energy assumes the simpler form $F(T,...,\varepsilon^{(1)}.) = -\left(\frac{k_B T}{8d^3}\right) f(d, \mu^{(0)}, \mu^{(1)}, \varepsilon^{(0)}, \varepsilon^{(1)})$, where

$$f(d,..,\varepsilon^{(0)}, \varepsilon^{(1)}) \approx \sum_{\substack{\alpha,\beta \\ \alpha \neq \beta}} \int_0^\infty \kappa^2 d\kappa \, e^{-\kappa} \left\{\eta_\alpha(0) \Pi_\beta^{(n=0 \to n'=1),(n=1 \to n'=2)}\left(0, \frac{\kappa}{2d}\right)\right\}, \qquad (11)$$

$$\Pi_m^{(n=0 \to n'=1),(n=1 \to n'=2)}\left(0, \frac{\kappa}{2d}\right) = \frac{\left[\left(\frac{\varepsilon^{(1)} - \varepsilon^{(0)}}{\varepsilon^{(1)} + \varepsilon^{(0)}}\right) + \left(\frac{\varepsilon^{(2)} - \varepsilon^{(1)}}{\varepsilon^{(2)} + \varepsilon^{(1)}}\right) e^{-\frac{\kappa D}{d}}\right]}{\left[1 + \left(\frac{\varepsilon^{(1)} - \varepsilon^{(0)}}{\varepsilon^{(1)} + \varepsilon^{(0)}}\right)\left(\frac{\varepsilon^{(2)} - \varepsilon^{(1)}}{\varepsilon^{(2)} + \varepsilon^{(1)}}\right) e^{-\frac{\kappa D}{d}}\right]}, \qquad (12)$$

$$\Pi_e^{(n=0 \to n'=1),(n=1 \to n'=2)}\left(0, \frac{\kappa}{2d}\right) = \frac{\left[\left(\frac{\mu^{(1)} - \mu^{(0)}}{\varepsilon^{(1)} + \varepsilon^{(0)}}\right) + \left(\frac{\mu^{(2)} - \mu^{(1)}}{\mu^{(2)} + \mu^{(1)}}\right) e^{-\frac{\kappa D}{d}}\right]}{\left[1 + \left(\frac{\mu^{(1)} - \mu^{(0)}}{\mu^{(1)} + \mu^{(0)}}\right)\left(\frac{\mu^{(2)} - \mu^{(1)}}{\mu^{(2)} + \mu^{(1)}}\right) e^{-\frac{\kappa D}{d}}\right]}. \qquad (13)$$

The replacements $\mu^{(1)} \to Z^{(1)^2}\varepsilon^{(1)}$, $\eta_m(0) \to r(0)^2 \eta_e(0)$, etc., for the analysis purpose, enable us to write the Casimir-Polder force as

$$\acute{K}(d,T,Z^{(1)},r(0),\eta_e(0),\varepsilon^{(1)}) = -3\left(k_B T \frac{f(d,Z^{(1)},r(0)\eta_e(0),\varepsilon^{(1)})}{8D^4}\right)\left(\frac{D}{d}\right)^4$$

$$+ \left(k_B T \frac{d f''(d, Z^{(1)}, r(0), \eta_e(0), \varepsilon^{(1)})}{8D^4}\right) \left(\frac{D}{d}\right)^4. \qquad (14)$$

This is the formal expression of the Casimir-Polder force corresponding to interactions of an electrically and magnetically polarizable microparticle with a magneto-dielectric sheet. Here the impedance of the sheet is $\mathbf{Z^{(1)}} = \sqrt{(\mu^{(1)}/\varepsilon^{(1)})}$ where $(\mu^{(1)}, \varepsilon^{(1)})$ are the the magnetic permeability and the dielectric permittivity of the film material, respectively. Similarly, the polarizability ratio of the micro-particle in vacuum is $\mathbf{r(0)} = \sqrt{(\eta_m(0)/\eta_e(0))}$ where $\eta_m(0)$ and $\eta_e(0)$, respectively, are the static magnetic and the electric polarizability of the micro-particle in vacuum. Note that the Casimir/ Casimir-Polder force arises from fluctuations of the electromagnetic field which are purely quantum-mechanical objects. At nonzero temperatures, the fluctuations also have a thermal contribution **[1,18]**. In our approximation of ignoring the frequency dependence completely, a significant physical information is lost: The formula (9) is written in terms of the imaginary frequencies though it has a representation in the real frequency domain as well **[19,20]**. The latter enables one to analyze the contributions from propagating and evanescent waves separately. At small distance the repulsive evanescent contributions are found to be dominating for the transverse electric polarization in the case of metallic objects**[20]**. Therefore, the thermal contributions need to be taken into account to discuss the Casimir-Polder repulsion. Furthermore, if the film happens to be isolated, then $\varepsilon^{(2)}(0)$ and $\mu^{(2)}(0) = 1$. All these restrictions enable us to write

$$\mathfrak{I}_m^{(n=0, n'=1)}(\varepsilon^{(1)}) = -\mathfrak{I}_m^{(n=1, n'=0)}(\varepsilon^{(1)}) = \left(\frac{\varepsilon^{(1)} - 1}{\varepsilon^{(1)} + 1}\right),$$

$$\mathfrak{I}_e^{(n=0, n'=1)}(\varepsilon^{(1)}, Z^{(1)}) = -\mathfrak{I}_e^{(n=1, n'=0)}(\varepsilon^{(1)}, Z^{(1)}) = \left(\frac{\mu^{(1)} - \mu^{(0)}}{\mu^{(1)} + \mu^{(0)}}\right) = \left(\frac{\varepsilon^{(1)} Z^{(1)2} - 1}{\varepsilon^{(1)} Z^{(1)2} + 1}\right). \qquad (15)$$

We can obtain, in principle, the Casimir-Polder energy by evaluating the integral(11) considering the terms in the integrand when $D/d \ll 1$. The limit $D/d \gg 1$ does not make sense. In the limit $D/d \ll 1$, to the leading order, the Casimir-Polder force is given by

$$Ḱ(d, T, Z^{(1)}, r(0)) = -(3k_B T\, Д(\varepsilon^{(1)}, Z^{(1)}, r(0))D/4d^5),$$

$$Д(\varepsilon^{(1)}, Z^{(1)}, r(0)) = (\eta_e(0)\, \varepsilon^{(1)})\{(1 + Z^{(1)2}\, r(0)^2) - (Z^{(1)2} + r(0)^2)/(\varepsilon^{(1)2} Z^{(1)2})\}. \qquad (16)$$

This is the Casimir-Polder force in the relatively large-separation limit. For dielectric film with no magnetic properties, the Casimir-Polder free energy and force are obtained from expressions for $F(d,T)$ and $Ḱ(d,T)$, respectively, by putting $\mu_0^{(1)} = 1$. This force is generally attractive.

The attractive nature in the large-separation limit is expected as the force has deep connection with the van der Waals force. To explain, we first note the well-known **[10,11,12,13]** fact that the atoms and molecules acquire temporary dipole moments as the space between interacting bodies is inhabited by "virtual" particle–antiparticle pairs fated to get annihilated in the time $\sim (\Delta E)^{-1}$ where $\Delta E$ is the energy of the fluctuations. These particle–antiparticle pairs, acting as the dipoles, radiate electromagnetic fields outward, and the fields interact with and scatter from fluctuating dipoles in the same as well as other macroscopic bodies. The reason for the scattering from the fluctuating dipoles being that the radiated fields can propagate over considerably long spatial ranges. To explain further, we take the convenient example of two parallel conducting plates, packed with radiating dipoles, with radiations of longer wavelengths dominating generally. Now the ability of dipoles on one plate to radiate at long wavelengths to dipoles on the other plate compared to the directions away from the plates is severely jeopardized due to the dipole moment and conductivity related boundary conditions. As a result of which the radiations away from the plates will be of much larger intensity compared to those directed towards the other plate. In other words, the long-wavelength radiation has more opportunities to leave the pair of plates entirely than to go between the plates, Inevitably, the impulse opposite to the former would push each plate toward the other. Having explained why the force is generally attractive, we hope that the explanation and calculation above set the tenor of the discussions to follow. It will be relevant to add that, for conducting parallel flat plates separated by a distance $d$, this force per unit area has the magnitude $(\pi^2/240)(\frac{\hbar c}{d^4})$ as calculated in ref.[1]. The role of $c$ above is to convert the electromagnetic mode wavelength to a frequency, while $\hbar$ converts the frequency to an energy. The absence of electronic charge implies that the electromagnetic field does not couple to matter in the usual sense here. It is important to note that when the separation, $d$, is so small that the mode frequencies are higher than the plasma frequency (for a metal) or higher than the surface Plasmon resonances or SPR (for a dielectric) of the material used to make the plates, this result breaks down.

To discuss the limit $D/d \sim 1$, we expand $e^{-\frac{\kappa D}{d}}$ in (11). In view of this and Eq. (15), the Casimir-Polder free energy assumes the simple form $F(d,,...T) = -\left(\frac{k_B T}{8d^3}\right) \sum_{\substack{\alpha,\beta \\ \alpha \neq \beta}} \wp_{\alpha\beta}\left(\varepsilon^{(1)}, Z^{(1)}, \eta_m(0), \frac{D}{d}\right)$, where the summand is

$$\wp_{\alpha\beta}\left(\varepsilon^{(1)}, Z^{(1)}, r(0), \frac{D}{d}\right) = \int_0^\infty \kappa^2 d\kappa\, e^{-\kappa} \left\{ \frac{\eta_\alpha(0)\Im_\beta^{(n=0,n'=1)}(\varepsilon^{(1)}, Z^{(1)})\left(1-e^{-\frac{\kappa D}{d}}\right)}{\left(1-\Im_\beta^{(n=0,n'=1)^2}(\varepsilon^{(1)}, Z^{(1)})\, e^{-\frac{\kappa D}{d}}\right)} \right\},$$

$$e^{-\frac{\kappa D}{d}} = Lim_{n\to\infty} \sum_{r=0}^{r=\infty} \frac{(-1)^r}{r!}\left\{1.\left(1-\frac{1}{n}\right)\left(1-\frac{2}{n}\right)........\left(1-\frac{r-1}{n}\right)\right\}\left(\frac{\kappa D}{d}\right)^r. \tag{17}$$

We have put in place all the relevant results. The repulsive forces arise in magnetic materials with non-trivial response functions **[21,22]**. As it has already been mentioned that they also arise for fluid-separated geometries **[4],** magneto-electric materials **[5, 23]**, and the TRS broken materials **[10,11,12,13].** The Casimir-Polder free energy $F(d, T, Z^{(1)}, ....)$ could also be written as $F(d, T, Z^{(1)}, ....) = -\left(\frac{k_B T}{8d^3}\right) f(d, Z^{(1)}, \eta_m(0), ...)$, where

$$f(d, \varepsilon^{(1)}, Z^{(1)}, \eta_m(0), ...) = \sum_{\substack{\alpha,\beta \\ \alpha \neq \beta}} \int_0^\infty \Sigma_l \eta_\alpha(0) Л_\beta^{(0)}(..., \kappa, d, Z^{(1)}, ...) \kappa^2\, e^{-\kappa}\, d\kappa,$$

$$Л_\beta^{(0)} = \frac{\Im_\beta^{(0)}\left(1-e^{-\frac{\kappa D}{d}}\right)}{\left(1-\Im_\beta^{(0)^2} e^{-\frac{\kappa D}{d}}\right)}, \tag{18}$$

the superscript '(0)' stands for the zero frequency limit. The quantities $\Im_\beta^{(0)}$ are given by Eq.(15). The term-by-term integration of

$$f\left(d, \varepsilon^{(1)}, Z^{(1)}, ...\right) = \sum_{\substack{\alpha,\beta \\ \alpha \neq \beta}} \int_0^\infty \Sigma_l \eta_\alpha(0) Л_\beta^{(0)}(..., \kappa, d, Z^{(1)}, ...) \kappa^2\, e^{-\kappa}\, d\kappa$$

and a little algebra, eventually yield $F(d, T, Z^{(1)}, \eta_m(0), ...) = -\left(\frac{k_B T}{8d^3}\right) f(d, Z^{(1)}, \eta_m(0), ...)$, where

$$f(d, Z^{(1)}, \eta_m(0), ...) = \sum_{m=1}^{\infty}(-1)^{m-1} \sum_{\substack{\alpha,\beta \\ \alpha \neq \beta}} \frac{\eta_\alpha(0) \mathfrak{I}_\beta^{(n=0,n'=1)}(\varepsilon^{(1)}, Z^{(1)})}{\left(1 - \mathfrak{I}_\beta^{(n=0,n'=1)^2}(\varepsilon^{(1)}, Z^{(1)})\right)} [(m^2 + 3m + 2)$$

$$\times \left(\frac{D}{d}\right)^m \times \{1 + (2^m - 2) \frac{\mathfrak{I}_\beta^{(n=0,n'=1)^2}(\varepsilon^{(1)}, Z^{(1)})}{(1 - \mathfrak{I}_\beta^{(n=0,n'=1)^2}(\varepsilon^{(1)}, Z^{(1)}))} \}]. \quad (19)$$

It may be noted that the contributions to $\kappa$ − integration in, say, Eq.(19) arises from the exponential terms $e^{-\frac{\kappa D}{d}}$, and not from $\mathfrak{I}_\beta^{(n=0,n'=1)}(\varepsilon^{(1)}, Z^{(1)})$. So, the term-by-term integrations are not very cumbersome. Eq.(19) immediately yields the Casimir-Polder force as $\check{K}(d, T, \varepsilon^{(1)}, Z^{(1)}, r(0)) = -(k_B T/8d^4) \, g(d, \varepsilon^{(1)}, Z^{(1)}, r(0))$ where

$$g(d, \varepsilon^{(1)}, Z^{(1)}, r(0)) = [3 f(d, \varepsilon^{(1)}, Z^{(1)}, r(0)) - d \, f'(d, \varepsilon^{(1)}, Z^{(1)}, r(0))]$$

$$= \sum_{\substack{\alpha,\beta \\ \alpha \neq \beta}} \sum_{m=1}^{\infty}(-1)^{m-1} \frac{\eta_\alpha(0) \mathfrak{I}_\beta^{(n=0,n'=1)}(\varepsilon^{(1)}, Z^{(1)})}{\left(1 - \mathfrak{I}_\beta^{(n=0,n'=1)^2}(\varepsilon^{(1)}, Z^{(1)})\right)} [(m^3 + 6m^2 + 11m + 6)$$

$$\times \{1 + (2^m - 2) \frac{\mathfrak{I}_\beta^{(n=0,n'=1)^2}(\varepsilon^{(1)}, Z^{(1)})}{(1 - \mathfrak{I}_\beta^{(n=0,n'=1)^2}(\varepsilon^{(1)}, Z^{(1)},))} \} \times \left(\frac{D}{d}\right)^m]. \quad (20)$$

Upon expanding (20), a little rearrangement of terms enables us to write

$$g(d, \varepsilon^{(1)}, Z^{(1)}, r(0))$$

$$= -24\eta_e(0) \left(\frac{D}{d}\right) \times [5\left(\frac{D}{d}\right)(\mathfrak{I}_m a_e^2 + \mathfrak{I}_e a_m^2 r(0)^2) \tilde{I}_1\left(\frac{D}{d}\right) - (a_e + a_m r(0)^2) \tilde{I}_2\left(\frac{D}{d}\right)]$$

$$= 120\eta_e(0) \left(\frac{D}{d}\right) \tilde{I}_2\left(\frac{D}{d}\right)(\mathfrak{I}_m a_e^2 + \mathfrak{I}_e a_m^2 r(0)^2) \times \left\{ p - \left(\frac{\left(\frac{D}{d}\right)\tilde{I}_1\left(\frac{D}{d}\right)}{\tilde{I}_2\left(\frac{D}{d}\right)}\right) \right\}, \quad (21a)$$

$$p \equiv \left(\frac{(a_e + a_m r(0)^2)}{5(\mathfrak{I}_m a_e^2 + \mathfrak{I}_e a_m^2 r(0)^2)}\right), \quad (21b)$$

where $r(0) = \sqrt{(\eta_m(0)/\eta_e(0))}$ and $\tilde{I}_1\left(\frac{D}{d}\right)$ and $\tilde{I}_2\left(\frac{D}{d}\right)$ are the two slowly convergent series:

$$\tilde{I}_1\left(\frac{D}{d}\right) = [1 - 6\left(\frac{D}{d}\right) + \left(\frac{49}{2}\right)\left(\frac{D}{d}\right)^2 - \cdots],$$

$$\tilde{I}_2\left(\frac{D}{d}\right) = [1 - \left(\frac{5}{2}\right)\left(\frac{D}{d}\right) + 5\left(\frac{D}{d}\right)^2 - \left(\frac{35}{4}\right)\left(\frac{D}{d}\right)^3 + \cdots].$$

The other quantities, viz. $(\mathfrak{I}_m(\varepsilon^{(1)}), \mathfrak{I}_e(\varepsilon^{(1)}, Z^{(1)}))$, are defined in Eq.(15):

$$\mathfrak{I}_e^{(n=0,n'=1)}(\varepsilon^{(1)}, Z^{(1)}) = \left(\frac{\varepsilon^{(1)} Z^{(1)^2} - 1}{\varepsilon^{(1)} Z^{(1)^2} + 1}\right), \mathfrak{I}_m^{(n=0,n'=1)}(\varepsilon^{(1)}) = \left(\frac{\varepsilon^{(1)} - 1}{\varepsilon^{(1)} + 1}\right). \quad (22)$$

The undefined ones are $(a_e, a_m)$. These are given by $a_e(\varepsilon^{(1)}) = \mathfrak{I}_m(\varepsilon^{(1)})/(1 - \mathfrak{I}_m^2(\varepsilon^{(1)}))$ and $a_m(\varepsilon^{(1)}, Z^{(1)}) = \mathfrak{I}_e(\varepsilon^{(1)}, Z^{(1)})/(1 - \mathfrak{I}_e^2(\varepsilon^{(1)},))$. One immediately obtains a criterion for the attractive Casimir-Polder interaction to turn repulsive. As long as we have

$$\left(\frac{\left(\frac{D}{d}\right)\tilde{I}_1\left(\frac{D}{d}\right)}{\tilde{I}_2\left(\frac{D}{d}\right)}\right) < p, \tag{23a}$$

the function $g(d, \varepsilon^{(1)}, Z^{(1)}, r(0))$ is greater than zero, and therefore the force is attractive. When

$$\left(\frac{\left(\frac{D}{d}\right)\tilde{I}_1\left(\frac{D}{d}\right)}{\tilde{I}_2\left(\frac{D}{d}\right)}\right) > p, \tag{23b}$$

the force turns repulsive as $g(d, \varepsilon^{(1)}, Z^{(1)}, r(0)) < 0$. Our calculation above pertains to the Faraday-Maxwell (static) limit, where the frequency dependence of all functions are ignored completely, resulting in the appearence of the conditions (23) above. Furthermore, it is being hoped, notwithstanding the fact that the in-depth investigation of the present problem requires dealing with a quantum-mechanical description, that our semi-phenomeological approach with the Stoner-like criterion enshrined in (23) for the attraction-repulsion crossover, will generate interest among the Casimir Physics community to cast a fresh look to the problem. In fact, recently a methodology[24] dealing with the reversible contollability aspect of the Casimir effect has been reported. The underlying physical mechanism is that the external driving electric fields suppress the charge correlations which are responsible for the fluctuation interaction. We shall now investigate the Casimir-Polder interaction in the high-temperature limit. We shall also take the Matsubara frequency $\omega_l = 2\pi l k_B T/\hbar$ dependence in an approximate manner at a comparatively lower temperature.

In the high temperature (or, the large seperation) regime, the long time behavior of the ubiquitous dissipation is dominated by an exponential decay with a time constant given by the first Matsubara frequency $\omega_1 = 2\pi k_B T/\hbar$. At a given temperature $'T'$, for the micro-particle–sheet system, it is then desirable that $\omega_1 \gg \frac{c}{d}$ for the large seperation limit and $\omega_1 \gtrsim \frac{c}{d}$, but not much higher, for comparatively moderate separations. This simply implies that $T >> T_c \equiv \hbar c/(2\pi d k_B)$ and $T \gtrsim T_c$, respectively, for the former and the latter cases. The important outcome, unlike the Faraday-Maxwell (static) limit, is that in the high-temperature limit $k^{(n)}(i\omega_l, k^\perp) \approx k^{(n)}(i\omega_l, 0) = [\varepsilon^{(n)}(i\omega_l) \mu^{(n)}(i\omega_l)(\omega_l^2/c^2)]^{1/2} \approx (\varepsilon^{(n)}(0) \mu^{(n)}(0))^{1/2}(\omega_l/c)$, and for the comparatively moderate temperatures

$$k^{(n)}(i\omega_l, k^\perp) \approx (\varepsilon^{(n)}(0) \mu^{(n)}(0))^{1/2}(\omega_l/c)[1+ (k^{\perp 2}/(\varepsilon^{(n)}(0) \mu^{(n)}(0) (\omega_l/c)^2)]^{1/2}$$

$$\approx (\varepsilon^{(n)}(0) \mu^{(n)}(0))^{1/2} (\omega_l/c) [1+ (k^{\perp 2}/2(\varepsilon^{(n)}(0) \mu^{(n)}(0) (\omega_l/c)^2)]. \tag{24}$$

It is easy to see that, for the former case, the criterion for the attractive Casimir-Polder interaction to turn repulsive is formally given by Eq.(23). Only the quantities, such as $(\mathfrak{I}_m(\varepsilon^{(1)}), \mathfrak{I}_e(\varepsilon^{(1)}, Z^{(1)}))$, are not defined anymore by Eq.(22). These are rather given by $\mathfrak{I}_e^{hTl}(\varepsilon^{(1)}, Z^{(1)}) \approx \left(\frac{\varepsilon^{(1)2} Z^{(1)3} - 1}{\varepsilon^{(1)2} Z^{(1)3} + 1}\right)$ and $\mathfrak{I}_m^{hTl}(\varepsilon^{(1)}, Z^{(1)}) \approx \left(\frac{\varepsilon^{(1)2} Z^{(1)} - 1}{\varepsilon^{(1)2} Z^{(1)} + 1}\right)$; $(\varepsilon^{(0)}, \mu^{(0)})$ have been put to one above. The superscript $'hTl'$ stands for the high-temperature limit. Strictly speaking, we shall have to consider the frequency ($\omega$) dependences of the permeability ($\mu$) and the permittivity ($\varepsilon$) of the sheet material as well, for all information about the optical properties of the surface is encoded in these response functions. In the final leg of this article, we shall take up this issue. We emphasize that, as long as the frequency dependence of the response functions are ignored, the crucial result, when the frequency dependence is completely ignored, given by Eq. (23) is not formally different from the high-temperature limit result.

For the not-so-high temperatures case, we shall have

$$\mathfrak{I}_e^{(n=0,n'=1)}\left(\kappa,T,d,\varepsilon^{(1)},Z^{(1)}\right)=\left(\frac{\varepsilon^{(1)2}Z^{(1)3}-\sqrt{(1+\frac{\kappa^2}{a_0^2(T)})}/\sqrt{(1+\frac{\kappa^2}{a_1^2(T)})}}{\varepsilon^{(1)2}Z^{(1)3}+\sqrt{(1+\frac{\kappa^2}{a_0^2(T)})}/\sqrt{(1+\frac{\kappa^2}{a_1^2(T)})}}\right), a_0^2(T)=\left(\frac{D}{d}\right)^{-2}\left(\frac{2D\omega_1}{c}\right)^2, (25)$$

$$\mathfrak{I}_m^{(n=0,n'=1)}\left(\kappa,T,d,\varepsilon^{(1)},Z^{(1)}\right)=\left(\frac{\varepsilon^{(1)2}Z^{(1)}-\sqrt{(1+\frac{\kappa^2}{a_0^2(T)})}/\sqrt{(1+\frac{\kappa^2}{a_1^2(T)})}}{\varepsilon^{(1)2}Z^{(1)}+\sqrt{(1+\frac{\kappa^2}{a_0^2(T)})}/\sqrt{(1+\frac{\kappa^2}{a_1^2(T)})}}\right), a_1^2(T)=n^{(1)2}\left(\frac{D}{d}\right)^{-2}\left(\frac{2D\omega_1}{c}\right)^2,$$

(26)

where $n^{(1)2}=\left(\varepsilon^{(1)}\mu^{(1)}\right)=\varepsilon^{(1)2}Z^{(1)2}$. The temperature lowering can lead to a complete cancellation or the change of sign of the micro-particle–sheet interaction. We note that now the contributions to $\kappa$ − integration in, say, Eq.(17) arise from the exponential term $e^{-\frac{\kappa D}{d}}$, as well as from $\mathfrak{I}_\beta^{(n=0,n'=1)}\left(\kappa,T,d,\varepsilon^{(1)},Z^{(1)}\right)$. So, the term-by-term integrations and overall calculations will be a little more cumbersome. We wish to continue below together with the expansions, of $(\mathfrak{I}_e,\mathfrak{I}_m)$ in (25) and (26), under the moderately high temperature assumption reflected in the inequality $\frac{\kappa^2}{a_0^2(T)}\ll 1$. The problem is tractable under this assumption. The expansions are

$$\mathfrak{I}_e^{(n=0,n'=1)}\left(\kappa,T,d,\varepsilon^{(1)},Z^{(1)}\right)\approx\left[\mathfrak{I}_e^{hTl}\left(\varepsilon^{(1)},Z^{(1)}\right)-A_e(\varepsilon^{(1)},Z^{(1)},T)\left(\frac{D}{d}\right)^2\kappa^2+O(\kappa^4/4a_0^2a_1^2)\right],$$

(27)

$$\mathfrak{I}_e^{hTl}\left(\varepsilon^{(1)},Z^{(1)}\right)\approx\left(\frac{\varepsilon^{(1)2}Z^{(1)3}-1}{\varepsilon^{(1)2}Z^{(1)3}+1}\right), A_e(\varepsilon^{(1)},Z^{(1)},T)=\left(\frac{\left(\frac{2D\omega_1}{c}\right)^{-2}(1-\varepsilon^{(1)-2}Z^{(1)-2})}{\left(\varepsilon^{(1)}Z^{(1)\frac{3}{2}}+\frac{1}{\varepsilon^{(1)}Z^{(1)\frac{3}{2}}}\right)^2}\right),$$

(28)

$$\mathfrak{I}_m^{(n=0,n'=1)}\left(\kappa,T,d,\varepsilon^{(1)},Z^{(1)}\right)\approx\left[\mathfrak{I}_m^{hTl}\left(\varepsilon^{(1)},Z^{(1)}\right)-A_m(\varepsilon^{(1)},Z^{(1)},T)\left(\frac{D}{d}\right)^2\kappa^2+O(\kappa^4/4a_0^2a_1^2)\right],$$

(29)

$$\mathfrak{I}_m^{hTl}\left(\varepsilon^{(1)},Z^{(1)}\right)\approx\left(\frac{\varepsilon^{(1)2}Z^{(1)}-1}{\varepsilon^{(1)2}Z^{(1)}+1}\right), A_m(\varepsilon^{(1)},Z^{(1)},T)=\left(\frac{\left(\frac{2D\omega_1}{c}\right)^{-2}(1-\varepsilon^{(1)-2}Z^{(1)-2})}{\left(\varepsilon^{(1)}Z^{(1)\frac{1}{2}}+\frac{1}{\varepsilon^{(1)}Z^{(1)\frac{1}{2}}}\right)^2}\right).$$

(30)

Furthermore, from Eqs. (7) and (8) we obtain

$$Л_\alpha(...,\kappa,d,\omega_l)=\left(\frac{\{\mathfrak{I}_\alpha^{hTl}(..,Z^{(1)})-A_\alpha\left(...,\omega_l,T\right)\left(\frac{D}{d}\right)^2\kappa^2+O\left(\frac{\kappa^4}{4a_0^2a_1^2}\right)\}\times\left(1-e^{-\frac{\kappa D}{d}}\right)}{\left[1-\{\mathfrak{I}_\alpha^{hTl2}(..,Z^{(1)})-2\mathfrak{I}_\alpha^{hTl}(..,Z^{(1)})A_\alpha(...,T)\left(\frac{D}{d}\right)^2\kappa^2+A_\alpha^2(...,T)\left(\frac{D}{d}\right)^4\kappa^4\}e^{-\frac{\kappa D}{d}}\right]}\right). \quad (31)$$

Using Eqs.(28)-(31), we find that for the comparatively moderate temperatures, the formal free energy expression could be written as

$$F(d,T,\varepsilon^{(1)},\ \mu^{(1)},\eta_e(0),\eta_m(0))=-\left(\frac{k_B T}{8d^3}\right)f(d,T,\varepsilon^{(1)},\ \mu^{(1)},\eta_e(0),\eta_m(0)), \tag{32}$$

where

$$f(.,d,T,\varepsilon^{(1)},..) = \sum_{\substack{\alpha,\beta \\ \alpha\neq\beta}} \int_0^\infty \sum_l \eta_\alpha \Lambda_\beta\left(\varepsilon^{(1)},Z^{(1)},\kappa,d,\omega_l\right) \kappa^2 e^{-\kappa} d\kappa,$$

$$\approx \sum_{\substack{\alpha,\beta \\ \alpha\neq\beta}} \int_0^\infty \sum_l \eta_\alpha \Lambda_\beta^{hTl} \kappa^2 e^{-\kappa} d\kappa$$

$$-\sum_\alpha \eta_\alpha(0) A_\alpha(..\omega_1,T)\left(\frac{D}{d}\right)^2 \frac{\left(\mathfrak{I}_\alpha^{hTl\,2}+1\right)}{\left(\mathfrak{I}_\alpha^{hTl\,2}-1\right)} \int_0^\infty \frac{c_\alpha(d_\alpha+1)}{(c_\alpha-1)^2}(\kappa^4+O(\kappa^6))e^{-\kappa} d\kappa, \tag{33}$$

We have used (17),(28)-(31) in Eq.(33). In the expansion in (33) we do not consider the terms of $O(\kappa^6)$ −order higher than what we have shown. The various terms in (33) are given by

$$\Lambda_\beta^{hTl} = \frac{\mathfrak{I}_\beta^{hTl}\left(1-e^{-\frac{\kappa D}{d}}\right)}{\left(1-\mathfrak{I}_\beta^{hTl\,2} e^{-\frac{\kappa D}{d}}\right)},\ c_\beta=\left(\frac{\left(e^{\frac{\kappa D}{d}}-1\right)}{\left(\mathfrak{I}_\beta^{hTl\,2}-1\right)}\right),\ d_\beta=\left(\frac{\left(e^{\frac{\kappa D}{d}}-1\right)}{\left(\mathfrak{I}_\beta^{hTl\,2}+1\right)}\right). \tag{34}$$

With $|c_\beta|,\ |d_\beta|<<1$, one may approximate the second term in (33) as

$$-\sum_\alpha \eta_\alpha(0) A_\alpha(..\omega_1,T)\left(\frac{D}{d}\right)^3 \frac{\left(\mathfrak{I}_\alpha^{hTl\,2}+1\right)}{\left(\mathfrak{I}_\alpha^{hTl\,2}-1\right)^2}\left[\int_0^\infty \left\{\kappa^5+\frac{1}{2}\kappa^6\left(\frac{D}{d}\right)\frac{\left(\mathfrak{I}_\alpha^{hTl\,2}+1\right)}{\left(\mathfrak{I}_\alpha^{hTl\,2}-1\right)}\right)+O(\kappa^7)\right\}e^{-\kappa}d\kappa\right], \tag{35}$$

Upon integration, the term within parentheses yields $120\left[1+3\left(\frac{D}{d}\right)\frac{\left(\mathfrak{I}_\beta^{hTl\,2}+1\right)}{\left(\mathfrak{I}_\beta^{hTl\,2}-1\right)}\right)+O\left(\left(\frac{D}{d}\right)^2\right)\right]$.

Since we are not considering the low temperature limit, we may replace $\omega_l$ by $\omega_1 = 2\pi k_B T/\hbar$ in the integral $f(..,d,\mu^{(0)},...)$ above. We then have

$$\mathfrak{I}_e^{hTl}\left(\varepsilon^{(1)},Z^{(1)}\right)\approx\left(\frac{\varepsilon^{(1)2}Z^{(1)3}-1}{\varepsilon^{(1)2}Z^{(1)3}+1}\right),\ A_e(\varepsilon^{(1)},Z^{(1)},T)=\left(\frac{\left(\frac{2D\omega_1}{c}\right)^{-2}\left(1-\varepsilon^{(1)-2}Z^{(1)-2}\right)}{\left(\varepsilon^{(1)}Z^{(1)\frac{3}{2}}+\frac{1}{\varepsilon^{(1)}Z^{(1)\frac{3}{2}}}\right)^2}\right), \tag{36}$$

$$\mathfrak{I}_m^{hTl}\left(\varepsilon^{(1)},Z^{(1)}\right)\approx\left(\frac{\varepsilon^{(1)2}Z^{(1)}-1}{\varepsilon^{(1)2}Z^{(1)}+1}\right),\ A_m(\varepsilon^{(1)},Z^{(1)},T)=\left(\frac{\left(\frac{2D\omega_1}{c}\right)^{-2}\left(1-\varepsilon^{(1)-2}Z^{(1)-2}\right)}{\left(\varepsilon^{(1)}Z^{(1)\frac{1}{2}}+\frac{1}{\varepsilon^{(1)}Z^{(1)\frac{1}{2}}}\right)^2}\right). \tag{37}$$

The first term in the series (33) above is $\sum_{\substack{\alpha,\beta \\ \alpha\neq\beta}} \int_0^\infty \sum_l \eta_\alpha \Lambda_\beta^{hTl} \kappa^2 e^{-\kappa} d\kappa$. This is given by Eq.(18) albeit with a slight difference in the definition of $(\mathfrak{I}^{hTl}{}_m,\ \mathfrak{I}^{hTl}{}_e)$. The functions $(\mathfrak{I}^{hTl}{}_m,\ \mathfrak{I}^{hTl}{}_e)$ are now given by (36) and (37). The second term in the series (33) above is given by (35). Thus the function $f(.,d,T,\varepsilon^{(1)},..)$ is given by

$$f(.,d,T,\varepsilon^{(1)},..) = \sum_{\substack{\alpha,\beta \\ \alpha\neq\beta}} \int_0^\infty \sum_l \eta_\alpha \Lambda_\beta\left(\varepsilon^{(1)},Z^{(1)},\kappa,d,\omega_1\right) \kappa^2 e^{-\kappa} d\kappa$$

$$=\sum_{m=1}^{\infty}(-1)^{m-1}\sum_{\substack{\alpha,\beta\\ \alpha\neq\beta}}\frac{\eta_\alpha(0)\Im_\beta^{hTl}(\varepsilon^{(1)},Z^{(1)})}{\left(1-\Im_\beta^{hTl2}(\varepsilon^{(1)},Z^{(1)})\right)}[(m^2+3m+2)$$

$$\times\left(\frac{D}{d}\right)^m\times\{1+(2^m-2)\frac{\Im_\beta^{hTl}(\varepsilon^{(1)},Z^{(1)})}{(1-\Im_\beta^{hTl}(\varepsilon^{(1)},Z^{(1)}))}\}]$$

$$-120\sum_\alpha\eta_\alpha(0)A_\alpha(..\omega_1,T)\left(\frac{D}{d}\right)^3\frac{\left(\Im_\alpha^{hTl2}+1\right)}{\left(\Im_\alpha^{hTl2}-1\right)^2}[1+3\left(\frac{D}{d}\right)\frac{\left(\Im_\alpha^{hTl2}+1\right)}{\left(\Im_\alpha^{hTl2}-1\right)})+O(\left(\frac{D}{d}\right)^2)]. \quad (38)$$

The significant difference between (19) and (38) is as follows: While all the cefficients in (19) are temperature independent, it is obvious from (38) that,though the coefficients of $\left(\frac{D}{d}\right)^m$ ($m=0,1,2$) will be temperature independent, the remaining ones will be temperature dependent due to the function $A_\beta(..\omega_1,T)$. Owing to the presence of the last term in Eq.(38), it is not difficult to see that the correction to $g(d,\varepsilon^{(1)},Z^{(1)},r(0))$ in Eq.(20), involving the same function, is

$$\Delta g\left(....,\varepsilon^{(1)},Z^{(1)},T,...\right)=360\sum_\alpha\eta_\alpha(0)A_\alpha(..\omega_1,T)[\left(\frac{D}{d}\right)^4\frac{\left(\Im_\alpha^{hTl2}+1\right)^2}{\left(\Im_\alpha^{hTl2}-1\right)^3}+O(\left(\frac{D}{d}\right)^5)]. \quad (39)$$

With this correction,as in (23), we immediately find that as long as we have

$$\left(\frac{\left(\frac{D}{d}\right)\tilde{I}_1\left(\frac{D}{d}\right)}{\tilde{I}_2\left(\frac{D}{d}\right)}\right)<\left(\frac{(a_e\tilde{I}_{2e}(T)+a_m\tilde{I}_{2m}(T)r^2(0))}{\frac{5}{2}(\Im_m a_e^2+\Im_e a_m^2 r^2(0))}\right), \quad (40)$$

the force is attractive. When

$$\left(\frac{\left(\frac{D}{d}\right)\tilde{I}_1\left(\frac{D}{d}\right)}{\tilde{I}_2\left(\frac{D}{d}\right)}\right)>\left(\frac{\left(a_e\tilde{I}_{2e}(T)+a_m\tilde{I}_{2m}(T)r^2(0)\right)}{\frac{5}{2}\left(\Im_m a_e^2+\Im_e a_m^2 r^2(0)\right)}\right), \quad (41)$$

the force turns repulsive. Inequations (40) and (41) reduce to (23), as they should, when $\left(\frac{2D\omega_1}{c}\right)^{-2}\ll 1$ (high-temperature limit). Here

$$\tilde{I}_{2\beta}(T)\approx[1+\left(\frac{15}{\tilde{I}_2\left(\frac{D}{d}\right)}\right)\left(\frac{A_\beta(..\omega_1,T)}{a_\beta(\varepsilon^{(1)},....)}\right)\frac{\left(\Im_\alpha^{hTl2}+1\right)^2}{\left(\Im_\alpha^{hTl2}-1\right)^3}], \quad (42)$$

$$a_e(\varepsilon^{(1)})=\Im_m(\varepsilon^{(1)})/(1-\Im_m^2(\varepsilon^{(1)})), a_m\left(\varepsilon^{(1)},Z^{(1)}\right)=\Im_e\left(\varepsilon^{(1)},Z^{(1)}\right)/(1-\Im_e^2(\varepsilon^{(1)},)). \quad (43)$$

We notice that the Casimir-Polder(CP) force not only arises from the reflection coefficients of the electromagnetic fluctuations on the sheet material plus substrate, there is also thermal contribution **[1,18].** The contribution is through the dependence on the Matsubara frequencies**.**

We shall do some graphics now to see what does inequation (23) convey. Analyzing the high temperature and the moderate temperature conterparts of (23) one may not gain probably a very different insight compared to what could be obtained from it. Therefore, the analysis of these results are not in the agenda at the moment. The term within the parenthesis in the right-hand-side of (21a) is $F(d, \varepsilon^{(1)}, Z^{(1)}, r(0)) \equiv \left\{ p\tilde{I}_2\left(\frac{D}{d}\right) - \left(\left(\frac{D}{d}\right)\tilde{I}_1\left(\frac{D}{d}\right)\right) \right\}$. Upon expanding upto the fifth order, we find that $F(d, \varepsilon^{(1)}, Z^{(1)}, r(0))$ is a quintic in $\left(\frac{D}{d}\right)$:

$$F(d, \varepsilon^{(1)}, Z^{(1)}, r(0), \omega = 0) =$$

$$-(21p + 260.4)\left(\frac{D}{d}\right)^5 + (14p + 84)\left(\frac{D}{d}\right)^4 - (8.75p + 24.5)\left(\frac{D}{d}\right)^3 + (5p + 6)\left(\frac{D}{d}\right)^2 - (2.5p + 1)\left(\frac{D}{d}\right) + p$$

(44)

Thus, the criterion (23) could now be expressed as if $F(d, \varepsilon^{(1)}, Z^{(1)}, r(0))$ is greater (less) than zero, the Casimir-Polder interaction is attractive(repulsive). To set the tone and the tenor of the discussion, we investigate the situation first with the aid of Eq.(16). We suppose that the film materials have access to non-trivial permeability and permittivity. The force could then be repulsive as well as we see below. We have plotted $-Д(\varepsilon^{(1)}, Z^{(1)}, r(0))$ as a function of $Z^{(1)}$ for $\varepsilon^{(1)} = 14$, and $r(0) = 0.20$(curve1), 0.30(curve2), 0.40(curve3), and 0.50(curve4) in Figure 2 (In Figure 2, $r(0)$ has been indicated by $Z_2$). The curves are recliner-shaped. The bend of the recliners, where $\acute{K}(d, T, Z^{(1)}, r(0)) = -(3k_BT/4D^4)(D/d)^5$, shifts towards left as $r(0)$ decreases. We find that the force is generally attractive except at values of $Z^{(1)}(\sim 0.05)$ and the polarizability ratio $r(0)$ ($\sim 0.40 - 0.50$). The values indicate that, if the micro-particle has higher polarizability ratio compared to the magnetic response and the electric response ratio of the sheet, the repulsion is accessible.

For the 2D graphics, we have assumed $\varepsilon^{(1)} = 14$, $Z^{(1)} = 0.5$ and 1.00, and $r(0) \equiv \sqrt{(\eta_m(0)/\eta_e(0))} = (01, 02, 03, 04)$. With $(\varepsilon^{(0)}, \mu^{(0)}, \varepsilon^{(2)}, \mu^{(2)}, \eta_e(0)) = 1$, we depict the crucial part of the Casimir-Polder interaction, viz. the function $F(d, \varepsilon^{(1)}, Z^{(1)}, r(0), \omega = 0)$ in the static limit. We have approximated it by a quintic in $\left(\frac{D}{d}\right)$. In figure $3(a)$ ($\varepsilon^{(1)} = 14$, $Z^{(1)} = 0.5$, and $r(0) = (01, 02, 03, 04)$), we find that interaction is attractive as long as $\left(\frac{D}{d}\right) \lesssim 0.2$, or, $d \gtrsim 5D$. For $\left(\frac{D}{d}\right) > 0.2$ (or, $d < 5D$), the interaction is repulsive, while in figure $3(b)$ ($\varepsilon^{(1)} = 14$, $Z^{(1)} = 1.00$, and $r(0) = (01, 02, 03, 04)$), we find that interaction is attractive as long as $\left(\frac{D}{d}\right) \lesssim 0.1$, or, $d \gtrsim 10D$. For $\left(\frac{D}{d}\right) > 0.1$ (or, $d < 10D$), the interaction is repulsive. The results ( repulsion at smaller separation $\left(\frac{d}{D}\right) \sim 1$ and the attraction at larger separation ) depicted in Figure 3 were expected as the Casimir-Polder/Casimir forces are very closely linked with the van der Waals' force. Additionally, we notice that, for the repulsion purpose, the sheet material is relatively high in the magnetic response (and the micro-particle has low magnetic polarizability). Generally, it is known **[25,26]** that for this purpose one requires a magnetic response strong enough to dominate the electric response of the material in a broad range of frequencies. Since this stringent condition is not met by any natural material, there has been a quest for an artificial material whose properties could be tailored in this direction. On a quick side note, the frequency (ω) dependences of the permeability (μ) and the permittivity (ε) of the sheet material must be taken into consideration, for all information about the optical properties of the surface **[27]** is encoded in these response functions.

We have developed here a quasi-phenomenological approach for the CP repulsion problem and obtained a Stoner-like criterion (for ferromagnetism) given by Eq.(23) for the attraction-repulsion crossover, notwithstanding the fact that the comprehensive investigation of the present problem requires dealing with a quantum-mechanical description. The graphical representations reveal that a strong magnetic response must dominate over the electric response of the material under investigation in a broad range of frequencies for the CP repulsion to become a reality. The prediction regarding artificial materials, such as the meta-materials **[25,26]** (MM) and the chiral meta-materials **[28,29]**(CMM), with tunable magneto-dielectric properties fuelled the hope of realizing the Casimir/CP repulsion and nano-levitation effect on demand in the second half of the last decade. The quest for the exotic material capable to deliver the Casimir/CP repulsion appeared to have been met with initial success. The existence of a repulsive Casimir force was found to depend upon the strength of the chirality($\sigma$ ) **[28,29]**. It must be mentioned that the MMs are basically made of nanostructures carefully fabricated to access a particular electromagnetic feature. For instance, the simultaneous occurance of the negative values for the permittivity and the permeability is the requirement that yields a 'left-handed' medium in which light propagates with opposite phase and energy velocities--a condition described by a negative refractive index in the electromagnetic domain. The CMMs, on the other hand, are separate class of MMs where the refractive index $n \neq \sqrt{(\mu_r \varepsilon_r)}$. The hope, however, was dashed as the very conjecture of accesssing the repulsive Casimir effect based on the CMMs was adjudged to be doubtful**[30].** The reason shown by the authors **[30]** is that the proposal is irreconcilable with the causality and the passivity of the meta-materials. This had perhaps pushed the investigation trail back to the initial step. The recent developments in nano-fabrication/ design procedure of MMs **[31,32]** with specially tailored magneto-electric properties, however, have resulted in the regeneration of hope in the field on investigation of dispersion forces in the presence of MMs.Future theoretical work should focus on a quantum-mechanical description of the micro-particle and the exotic material sheet system, compatible with the causality and the passivity of the material, to tune up the condition for the attraction-repulsion crossover.

**Figures and Captions:**

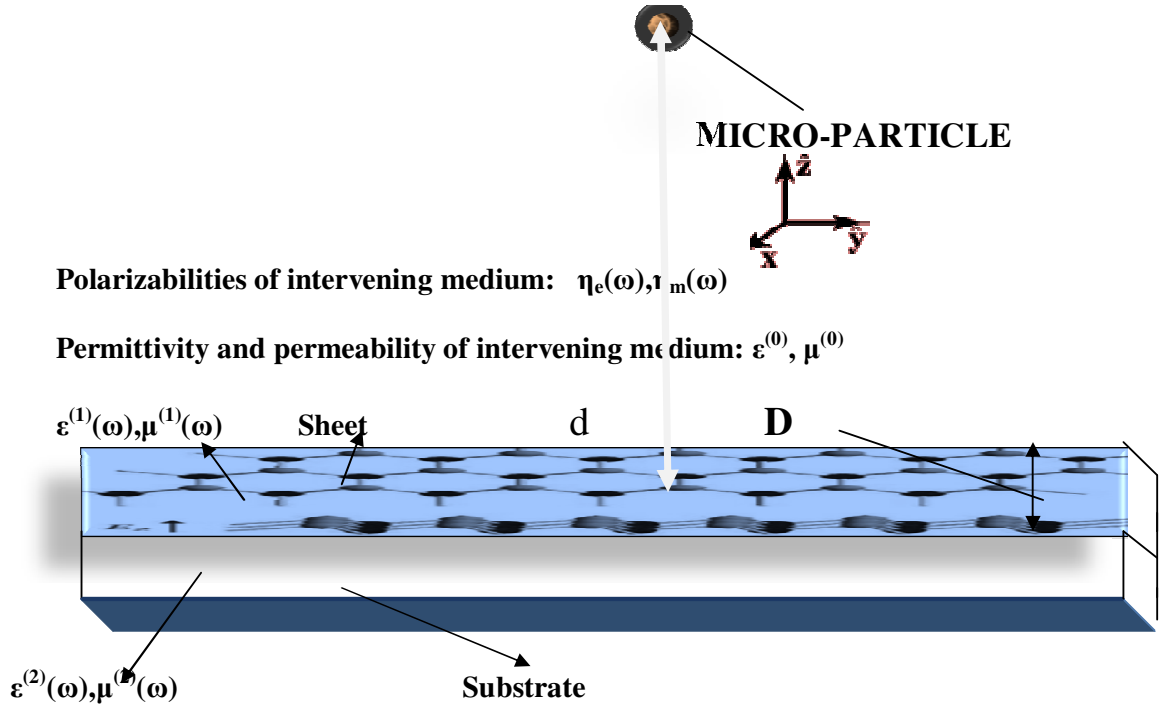

**Figure 1:** The configuration of a micro-particle in vacuum characterized by the electric polarizability $\eta_e(\omega)$ and the magnetic polarizability $\eta_m(\omega)$ at a distance 'd' above a sample consisting of thin sheet of thickness 'D' deposited on a thick substrate. While the sheet is characterized by the dielectric permittivity $\varepsilon^{(1)}(\omega)$ and the magnetic permeability $\mu^{(1)}(\omega)$, the substrate is by the permittivity $\varepsilon^{(2)}(\omega)$ and the permeability $\mu^{(2)}(\omega)$. We have chosen the coordinate plane (x, y) coinciding with the upper sheet surface and the z axis perpendicular to it.

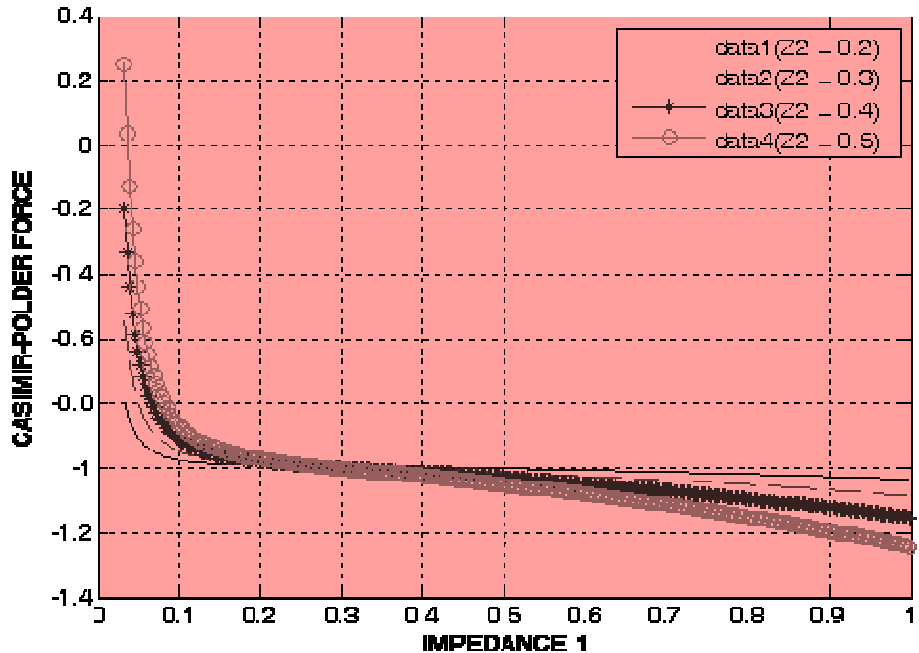

**Figure 2.** A plot of Casimir-Polder force in Eq.(16) as a function of $Z^{(1)}$ for $\varepsilon^{(1)} = 14$, and $r(0) = 0.20$(curve1), 0.30(curve2), 0.40(curve3), and 0.50(curve4) in the relatively large-separation limit. The force is generally attractive except at non-trivial values of $Z^{(1)} \sim 0.05$ and **r(0)** ~ 0.4-0.5.

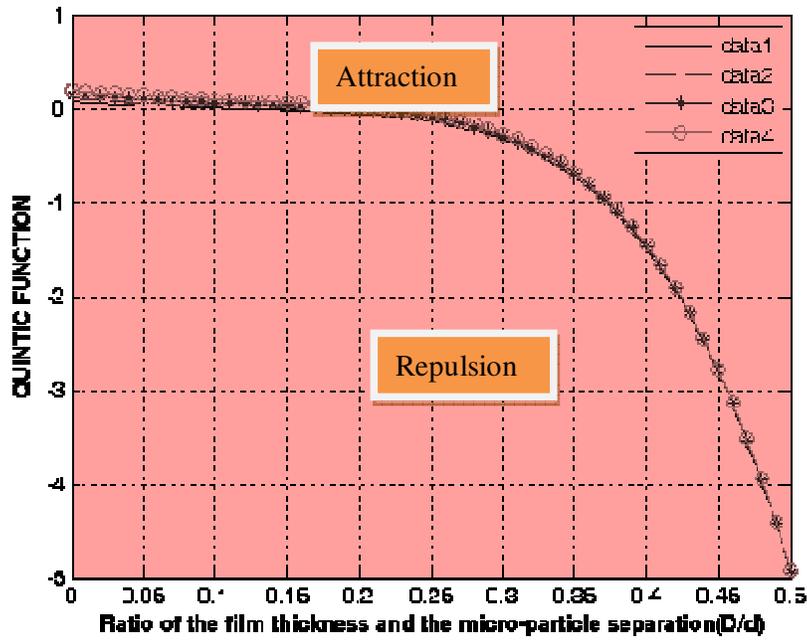

(a)

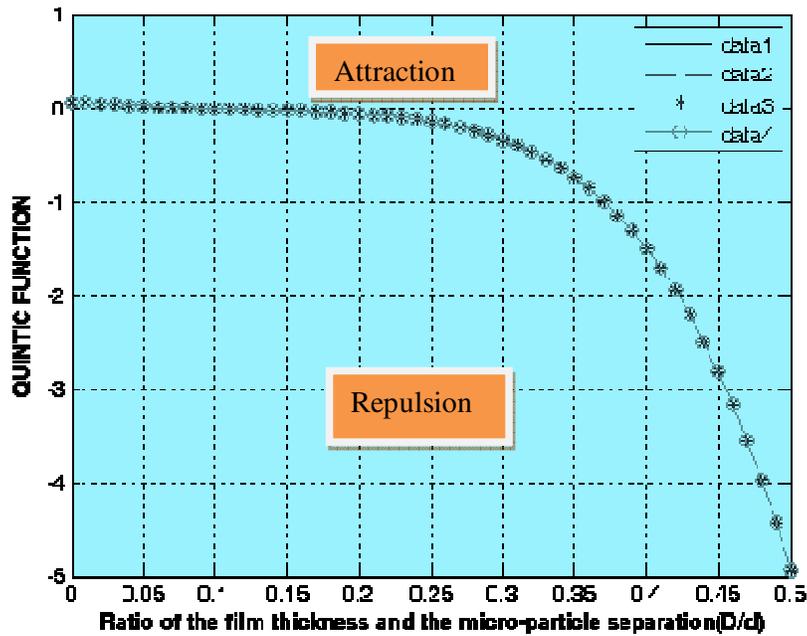

(b)

**Figure 3.** The **2D** plots of the quintic **F** *(d,...,r(0))* as a function of **(D/d)**; the remaining parameters are held fixed. **(a)** Here we have taken $\varepsilon^{(1)}$=14, $Z^{(1)}$=0.5  *r(0)* = **(01, 02, 03, 04)** .The quintic function is positive, i.e. the Casimir-Polder interaction is attractive as long as  (D/d) < **0.2**, or,  **d ≳ 5** . For (D/d) > **0.2**  **(or, d < 5D)**, the interaction is repulsive.**(b)** Here   $Z^{(1)}$= 1.00. The Casimir-Polder interaction is attractive as long as (D/d) < **0.1.**